\documentstyle[twocolumn,prl,aps,epsfig]{revtex}
\begin{document}                                                              
\newcommand{\be}{\begin{eqnarray}}
\newcommand{\ee}{\end{eqnarray}}
\newcommand{\dd}{\mathrm{d\,}}
\title{Quarkonium polarization in heavy ion collisions as \\
a possible signature of the quark--gluon plasma}
\author{B.L. Ioffe$^{a,b}$ and  D.E. Kharzeev$^b$}
\bigskip
\address{
\bigskip
a) Institute of Theoretical and Experimental Physics,\\
B. Cheremushkinskaya 25, Moscow 117218, Russia\\
\bigskip
b) Department of Physics,\\ Brookhaven National Laboratory,\\
Upton, New York 11973-5000, USA}
\date{June 6, 2003}
\maketitle
\begin{abstract} 
The polarization of quarkonium states produced in hadron collisions exhibits strong non--perturbative effects -- 
for example, at small transverse momentum $p_t$ charmonia appear unpolarized, in sharp contradiction to 
the predictions of perturbation 
theory.  
The quark--gluon plasma is expected to screen away the non--perturbative physics; therefore those quarkonia which 
escape from the plasma should possess polarization as predicted by perturbative QCD. We estimate 
the expected $J/\psi$ polarization at small $p_t$, and find that it translates into the asymmetry 
of the $e^+e^- (\mu^+\mu^-)$ angular distribution $W(\theta) \sim 1 + \alpha\ cos^2\theta$, 
with $\alpha \simeq 0.35 \div 0.4$.  
\end{abstract}
\begin{narrowtext}

The possibility to form quark--gluon plasma in heavy ion collisions is an intriguing 
problem of strong interaction physics.  To establish the formation of plasma, a number of 
signatures were proposed; here we will concentrate on heavy quarkonia. 
Suppression of heavy quarkonium states has been suggested long time ago by 
Matsui and Satz \cite{MS} as a signature 
of the deconfinement phase transition in heavy ion collisions. 
Their, by now well--known, idea is that the Debye screening of the gluon exchanges 
will make the binding of heavy quarks into the bound states impossible or unlikely once a sufficiently 
high temperature is reached. The lack of quarkonium states would thus 
signal deconfinement; this effect was indeed observed and studied in detail at CERN SPS by the 
NA38 \cite{NA38} and NA50 Collaborations \cite{NA50}. 
The results on $J/\psi$ production at RHIC have recently been presented 
by the PHENIX Collaboration \cite{phenix}. 
The observations of quarkonium suppression have been interpreted as a signal of 
quark--gluon plasma formation \cite{qgp}. However, different conclusions were reached in \cite{comov}, 
where it was argued that the effect may arise due to quarkonium collisions with the comoving 
hadrons. Additional tests of the quark--gluon plasma formation could help 
to clarify the situation. 

In this note we propose to use for the diagnostics of the quark--gluon plasma those 
heavy quarkonia which {\it escape} from it. This would require experimental measurements of 
quarkonium polarization, which can be reconstructed from the angular distributions of 
quarkonium decays -- dileptons and/or photons. For $J/\psi$ states, one would need to measure 
the angular distribution of electrons (or muons) in the $J/\psi \to e^+ e^-$ decay in $J/\psi$ rest frame 
relative to the direction of its momentum. (We will concentrate on  $J/\psi$'s at relatively small 
$p_t$, which dominate the total production cross section.)

\vskip0.3cm

Let us first formulate what we mean by the quark--gluon plasma, since different definitions 
sometimes may result in misunderstanding. We define the quark--gluon plasma as 
a gas of quarks and gluons in which the interactions can be described by perturbative QCD and 
non--perturbative effects are either absent or can be neglected. We will not need to specify 
the properties of this state of matter in more detail to develop our idea; let us now turn to 
the dynamics of quarkonium production.  

It is well--known that the description of the data on heavy quarkonium production 
within the framework of 
perturbative QCD (pQCD) meets with siginificant difficulties. Both the absolute values of the measured 
production 
cross sections of hidden heavy flavor states and the relative abundances of different quarkonia 
are not described well within the perturbative framework, but perhaps the most spectacular 
failure of pQCD is the polarization of the produced quarkonia. Even an extension of a perturbative 
approach based on non--relativistic QCD \cite{BBL}, which allows certain non--perturbative physics, 
does not allow to explain the polarization measurements \cite{CDF}.

Meanwhile, the description of 
heavy flavor production in perturbative framework has been largely successful (even though 
there are some problems there as well). The reason for this is easy to understand -- 
the production of heavy flavors occurs at short time scale $\sim 1/2m_Q$, where $m_Q$ is the 
heavy quark mass, whereas the binding of the produced heavy quarks into quarkonium is a softer 
process characterized by the time scale of $\tau_{bind} \sim 1/\epsilon$, where $\epsilon$ is the 
typical binding 
energy; for a Coulomb interaction, $\epsilon \sim \alpha_s m_Q^2 \ll m_Q$. The binding process 
is thus far more likely to be affected by non--perturbative phenomena, which manifest themselves 
both in the magnitude of the production cross section and in the polarization of the produced 
quarkonia. 

\vskip0.3cm

Consider now the production of quarkonium states in relativistic heavy ion collisions. The typical 
time scale for the production of semi--hard partons with transverse momentum $k_t$ is $\tau \sim 1/k_t$; 
for example, in the gluon saturation scenario $\tau_{prod} \sim 1/Q_s$, 
where $Q_s$ is the saturation scale which at RHIC energies is about 
$Q_s \simeq 1 \div 2$ GeV \cite{KLN}.  
It is thus likely that while these produced partons will not significantly affect the 
production of heavy quarks (which happens at earlier time), they will influence the binding 
of heavy quarks in quarkonia since $\tau_{prod} \le \tau_{bind}$. 

High energy density of the produced partonic state is expected to result in the destruction 
of the non--perturbative vacuum structure. Indeed, lattice QCD calculations show that quark and gluon 
condensate ``evaporate'' above the deconfinement phase transition \cite{lattice}. 
It may be expected that non--perturbative vacuum fluctuations are suppressed even if the 
thermalization does not take place -- a specific example is given by the suppression of instantons 
in the saturated gluon environment \cite{KKL}. As a result, the processes in this high--density 
partonic state of matter should be described by the weak coupling, perturbative methods. As a matter 
of fact, as we assumed above, 
one may {\it define} the quark--gluon plasma as a collective state of quarks and gluons the 
dynamics of which is governed by perturbative interactions. Therefore, the formation of heavy 
quarkonium states should also be adequately described by perturbation theory, and 
the predictions of pQCD for the polarization of heavy quarkonia should be vindicated. Dense parton matter 
may then screen out of existence a large part of quarkonia, as proposed  originally \cite{MS}, 
but those of them that survive will carry the information about the mechanism of their formation throughout  
the collision. Of course, the interactions of quarkonia at the later stages of the heavy ion collision 
may wash out their polarization somewhat, but relatively small interaction cross sections and the 
heavy quark symmetry, suppressing the spin flips of heavy quarks, should prevent quarkonia from 
``forgetting'' their initial polarization entirely. 
   
\vskip0.3cm

Let us illustrate this idea in more detail using the example of $J/\psi$ polarization. 
There are two mechanisms of $J/\psi$ production in hadron collisions -- direct, when $J/\psi$ is produced 
by perturbative and non--perturbative interactions of gluons and quarks, and cascade, when $J/\psi$ is 
created as a result of decays of C--even $\bar{c}c$ states, $\chi_c \to J/\psi + \gamma$. In quark--gluon 
plasma, the cascade production mechanism should be at least as important as direct production. 
Indeed, in the lowest order of perturbation 
theory,  $J/\psi$ is produced by the three gluon fusion or by two gluon fusion followed by the gluon 
emission off the $\bar{c}c$ system. In both cases the probability of   $J/\psi$ production is proportional 
to $\alpha_s^3(m_c)$. The probability of $\chi_c^{0,2}$ production is proportional to $\alpha_s^2(m_c)$, 
i.e. 
it is of lower order in $\alpha_s$, which however is largely compensated by
the branching ratio $B(\chi_2 \to J/\psi + \gamma) \simeq 20 \%$ for the $J/\psi$
production.

In hadron collisions the direct mechanism comprises typically about $60 \%$ of the observed  
 $J/\psi$'s (for a review of the data, see \cite{Gavai}), 
which seems to suggest that an essential part of 
 $J/\psi$ production in hadron collisions is of non--perturbative origin. Direct calculations confirm 
this conclusion. In ref \cite{VHBT}  $J/\psi$ production cross section in $\pi N$ interactions 
was calculated in perturbation theory: two gluon fusion into $\bar{c}c$ with the subsequent gluon 
emission (the so--called color--singlet model \cite{CSM}). The result is about $8$ times smaller than 
the data. Similar situation holds also for $\chi_2$ production: the calculated cross section is by factor 
of two smaller than the experimental one (see  \cite{VHBT} for details). Additional mechanism of 
$\chi_2$ production \cite{JK} in the framework of the color--octet model \cite{BBL} involves the 
formation of the color octet $\bar{c}c$ state which then decays by color $E1$ transition to $\chi_2$. 
Evidently, this mechanism perturbatively is suppressed by extra power of $\alpha_s$ and is essential 
only if it is non--perturbative. The cross section of $\chi_1$ production is very small in perturbation 
theory, but noticeable experimentally ($\chi_0$ does not contribute substantially to the 
$J/\psi$ production because of a small branching ratio of $\chi_0 \to J/\psi + \gamma$ decay -- about 
$1 \%$). (The contributions from various sources to the $J/\psi$ production in $\pi^- N$ collisions 
at the incident energy of $185$ and $300$ GeV and the results of 
theoretical calculations can be found in \cite{VHBT}; the comparison shows that the production 
of charmonium states in hadronic collisions is in an essential way non-perturbative).

Let us now turn to $J/\psi$ polarization as reconstructed from the angular distributions of electrons 
(muons) from the $J/\psi \to e^+ e^- (\mu^+ \mu^-)$ decays. Generally the electron (muon) distribution 
has the form
\be 
W(\theta) \sim 1 + \alpha \ cos^2 \theta, \label{dist}
\ee
where $\theta$ is the emission angle of $e^+$ (or $\mu^+$) relative to the direction of 
$J/\psi$ motion in its rest frame; at small $p_t$, this direction coincides with the direction 
of the beam. The value $\alpha = 1$ corresponds to the transverse polarization, $\alpha = -1$ -- 
to the longitudinal polarization, and $\alpha = 0$ to unpolarized $J/\psi$.  
In perturbation theory, in the case when  $J/\psi$ is produced through the 
  $\chi_2 \to J/\psi + \gamma$ decay, the coefficient $\alpha$ in Eq. (\ref{dist}) is determined 
unambiguously (at small $p_t$): $\alpha = 1$ \cite{BI}. This comes from the fact that $\chi_2$ is 
produced by two--gluon fusion, $gg \to \chi_2$, for which the effective interaction is 
$f_{\mu\nu} \Theta_{\mu\nu}$, where $\Theta_{\mu\nu}$ is the energy--mometum tensor of the gluon 
field and $f_{\mu\nu}$ is the wave function of $\chi_2$. Since $\Theta_{\mu\nu}$ has only $J_z = \pm 2$ 
spin projections on the direction of gluon momenta (indeed,  $\Theta_{\mu\nu}$ may be considered as a 
source of the graviton field), the same spin projections has the $\chi_2$. As a result, $J/\psi$ produced 
via $\chi_2$ decay is transversely polarized, $J_z = \pm 1$ and thus $\alpha =1$. 

This conclusion is somewhat modified when the initial transverse momenta of the gluons are taken 
into account. This reduces the value of $\alpha$ to \cite{BI} 
\be
\alpha \longrightarrow \alpha' = \alpha {(1 - {3 \over 2}\ \theta_0^2)\over 1 + \alpha\ \theta_0^2/2}, \label{pt}
\ee
where $\theta_0^2 \sim 4 \langle p_t^2 \rangle /M_{\chi}^2$. The average transverse momentum of 
gluons is expected to increase with energy and the atomic number of the colliding systems. For example,  
in the gluon saturation scenario $p_t \sim Q_s \sim A^{1/3} s^{\lambda/2}$, with $\lambda \simeq 0.25$; 
at RHIC energies in central $Au-Au$ collisions $Q_s \sim 1$ GeV \cite{KLN}. For $p_t \sim 1$ GeV, 
the formula Eq.(\ref{pt}) yields a reduction of polarization down to $\alpha \simeq 0.5$; still, 
this value corresponds to a significant transverse polarization.   

The asymmetry coefficient $\alpha$ was also computed for the directly produced $J/\psi$ and 
for the production via the $\chi_1$ decay \cite{VHBT}. The results are $\alpha_{dir} \simeq 0.25$ 
for direct production and $\alpha_{\chi_1} \simeq -0.15$ for the production via $\chi_1$ decay 
(except the forward region of $x_F > 0.8$, where both $\alpha_{dir}$ and $\alpha_{\chi_1}$ 
begin to increase). After summing all channels of $J/\psi$ production it was found \cite{VHBT} 
that $\alpha_{tot}^{pert} \simeq 0.5$. Experimentally \cite{data}, no sizable polarization 
in the entire range of $x_F$ was observed, $\alpha \simeq 0$ (there is however an indication 
that at very large $x_F$ $\alpha$ becomes negative). This disagreement between theory and experiment 
demonstrates again that the production mechanism of $J/\psi$ and possibly $\chi_1$ and $\chi_2$ 
in hadronic collisions is essentially non--perturbative. (Even though we have discussed only 
$\pi N$ data, there is no reason to believe that in $pN$ collisions the situation will be 
very different, apart from a relatively smaller contribution of the $\bar{q}q$ annihilation 
in the latter case.) It is also interesting to note that for the case of $\Upsilon$ production, 
the data from E866 Collaboration \cite{FNAL} show transverse polarization for $\Upsilon(2S+3S$, 
in qualitative agreement with the predictions of perturbation
theory. This of course is to be 
expected if the validity of perturbation theory were to improve between the scales fixed by the masses 
of charm and bottom quarks.         

\vskip0.3cm

Let us now dwell upon the $J/\psi$ production in heavy ion collisions. Let us assume that 
at sufficiently high collision energy the quark--gluon plasma is formed. Due to the arguments 
presented above, the formation of quarkonia will thus take place in the plasma. This will of course  
result in the suppression of the formation probability \cite{MS}; moreover, the presence of the 
plasma is likely to affect the excited states more significantly, and the contribution of the 
excited quarkonium states to the observed yield of $J/\psi$ will thus change, which also can result 
in the change of the $J/\psi$ polarization \cite{Gupta}. 
If quarkonium is produced in the plasma, 
the non--perturbative effects 
should be absent (or small), and we are left only with the perturbative mechanism. 
Then, according to \cite{VHBT}  
about one half of $J/\psi$'s will be produced directly and another one half via $\chi_2 \to 
J/\psi + \gamma$. (The approximate equality of these contributions stems from the fact 
that the extra power of $\alpha_s$ in the direct production cross section is compensated by a 
relatively small branching ratio -- about $20 \%$ -- of the $\chi_2 \to 
J/\psi + \gamma$ decay; note also that $\chi/J/\psi$ ratio has been found to be independent 
of the collision energy -- see \cite{Gavai}.) We thus expect that the asymmetry coefficient 
of the electron (muon) angular distribution in the  $J/\psi \to e^+e^-(\mu^+\mu^-)$ decay 
in the case of quark--gluon plasma formation will increase from zero to about (at $p_t = 0$) 
$\alpha \simeq 0.6$. The account of the initial transverse momentum distribution of gluons 
as discussed above reduces asymmetry coefficient to 
\be
\alpha \simeq 0.35 \div 0.4. \label{finres}
\ee
Still, we expect a remarkable increase in the asymmetry coefficient when going from 
hadron to heavy ion collisions. 

Of course, there are effects which may result in some 
decrease of $\alpha$ in comparison with (\ref{finres}), notably a more accurate account of the 
transverse momentum distributions of gluons and, as also discussed above, the interactions 
of $J/\psi$ with the constituents of hadronic and/or quark--gluon fireball. 
However, we do expect an increase of $J/\psi$ polarization in heavy ion collisions if the 
quark--gluon plasma is formed there.     

\vskip0.3cm

B.I. is thankful to Larry McLerran for the hospitality at Brookhaven National Laboratory, where this 
work was done. The work of B.I. was supported in part by INTAS Grant 2000--587 and RFBR Grant 03--02--16209. 
The research of D.K. is supported by the U.S. Department of Energy under Contract No. DE-AC02-98CH10886.

\end{narrowtext}
\end{document}